\documentclass[aps,twocolumn,showkeys,showpacs,preprintnumbers,prd,superscriptaddress,nofootinbib,10pt]{revtex4-1}
\usepackage{graphicx,epsf,bm,amsmath,amsfonts,amssymb,epstopdf,natbib,hyperref,color,verbatim,multirow,bm,appendix}
\hypersetup{colorlinks=true,urlcolor=blue,citecolor=blue,linkcolor=blue,menucolor=blue,anchorcolor=blue,filecolor=blue}
\date{\today}

\begin{document}

\title{Nonparametric late-time expansion history reconstruction and implications for the Hubble tension in light of recent DESI and type Ia supernovae data}

\author{Jun-Qian Jiang}
\email{jiangjunqian21@mails.ucas.ac.cn\\junqian.jiang@studenti.unitn.it}
\affiliation{School of Physical Sciences, University of Chinese Academy of Sciences, No.19(A) Yuquan Road, Beijing 100049, China}
\affiliation{Department of Physics, University of Trento, Via Sommarive 14, 38123 Povo (TN), Italy}

\author{Davide Pedrotti}
\email{davide.pedrotti-1@unitn.it}
\affiliation{Department of Physics, University of Trento, Via Sommarive 14, 38123 Povo (TN), Italy}
\affiliation{Trento Institute for Fundamental Physics and Applications (TIFPA)-INFN, Via Sommarive 14, 38123 Povo (TN), Italy}

\author{Simony Santos da Costa}
\email{simony.santosdacosta@unitn.it}
\affiliation{Department of Physics, University of Trento, Via Sommarive 14, 38123 Povo (TN), Italy}
\affiliation{Trento Institute for Fundamental Physics and Applications (TIFPA)-INFN, Via Sommarive 14, 38123 Povo (TN), Italy}

\author{Sunny Vagnozzi}
\email{sunny.vagnozzi@unitn.it}
\affiliation{Department of Physics, University of Trento, Via Sommarive 14, 38123 Povo (TN), Italy}
\affiliation{Trento Institute for Fundamental Physics and Applications (TIFPA)-INFN, Via Sommarive 14, 38123 Povo (TN), Italy}

\begin{abstract}
\noindent We nonparametrically reconstruct the late-time expansion history in light of the latest Baryon Acoustic Oscillation (BAO) measurements from DESI combined with various Type Ia Supernovae (SNeIa) catalogs, using interpolation through piece-wise natural cubic splines, and a reconstruction procedure based on Gaussian Processes (GPs). Applied to DESI BAO and PantheonPlus SNeIa data, both methods indicate that deviations from a reference $\Lambda$CDM model in the $z \lesssim 2$ unnormalized expansion rate $E(z)$ are constrained to be $\lesssim 10\%$, but also consistently identify two features in $E(z)$: a bump at $z \sim 0.5$, and a depression at $z \sim 0.9$, which cannot be simultaneously captured by a $w_0w_a$CDM fit. These features, which are stable against assumptions regarding spatial curvature, interpolation knots, and GP kernel, disappear if one adopts the older SDSS BAO measurements in place of DESI, and decrease in significance when replacing the PantheonPlus catalog with the Union3 and DESY5 ones. We infer $c/(r_dH_0)=29.90 \pm 0.33$, with $r_d$ the sound horizon at baryon drag and $H_0$ the Hubble constant. Breaking the $r_d$-$H_0$ degeneracy with the SH0ES prior on $H_0$, the significance of the tension between our nonparametric determination of $r_d=136.20^{+2.20}_{-2.40}\,{\text{Mpc}}$ and the \textit{Planck} $\Lambda$CDM-based determination is at the $5\sigma$ level, slightly lower than the $6\sigma$ obtained when adopting the older SDSS dataset in place of DESI. This indicates the persistence at very high significance of the ``sound horizon tension'', reinforcing the need for pre-recombination new physics. If substantiated in forthcoming data releases, our results tentatively point to oscillatory/nonmonotonic features in the shape of the expansion rate at $z \lesssim 2$, of potential interest for dark energy model-building.
\end{abstract}

\maketitle
\newpage

\section{Introduction}
\label{sec:introduction}

The nature of the dark energy (DE) component driving cosmic acceleration, initially discovered via observations of distant Type Ia Supernovae (SNeIa)~\cite{SupernovaSearchTeam:1998fmf,SupernovaCosmologyProject:1998vns} and now confirmed by a range of independent probes~\cite{Rubin:2016iqe,Nadathur:2020kvq,DiValentino:2020evt}, is arguably one of the biggest open problems in physics~\cite{Bamba:2012cp,Nojiri:2017ncd}. Regardless of whether DE will ultimately turn out to be the manifestation of a cosmological constant (CC), or of a dynamical, time-evolving component, unveiling its nature is likely to have enormous implications for several fields of physics. Over the past decades, significant focus has been placed on constraining DE's gravitational properties, as captured for instance by the redshift evolution of its energy density $\rho_x(z)$ or equation of state $w_x(z)$: some among the cleanest probes of these properties include background distance and expansion rate measurements, for instance through Baryon Acoustic Oscillations (BAO) in the clustering of tracers of the Large-Scale Structure (LSS) such as galaxies~\cite{Huterer:2017buf}.

The recent launch of various so-called ``Stage IV surveys'', among which the Dark Energy Spectroscopic Instrument (DESI) and Euclid, alongside the upcoming launch of several Cosmic Microwave Background (CMB) missions~\cite{SimonsObservatory:2018koc,SimonsObservatory:2019qwx,CMB-S4:2022ght}, are expected to be game-changers in the quest toward better understanding DE's (gravitational and non) properties. Such a pursuit is not only important per se, but has become rather pressing in recent years because of the emergence of persisting cosmological tensions, i.e.\ the disagreement between independent inferences of cosmological parameters. The most important among these is without doubt the Hubble tension~\cite{Verde:2019ivm,DiValentino:2021izs,Perivolaropoulos:2021jda,Schoneberg:2021qvd,Shah:2021onj,Abdalla:2022yfr,DiValentino:2022fjm,Hu:2023jqc,Vagnozzi:2023nrq,Verde:2023lmm} (and, to a lesser extent, the weaker tension in $S_8$~\cite{DiValentino:2018gcu,DiValentino:2020vvd,Nunes:2021ipq}), which could signal the breakdown of the concordance $\Lambda$CDM cosmological model~\cite{Akarsu:2024qiq}, and whose resolution may bear profound consequences for the nature of DE. However, such a quest has recently become even more pressing in light of the DESI Year 1 BAO results, which officially mark the start of the Stage IV DE era. The DESI results, especially once combined with a number of external probes, provide intriguing hints for a dynamical, time-evolving DE component~\cite{DESI:2024mwx,DESI:2024aqx,DESI:2024kob} -- it goes without saying that if substantiated, these hints may have tremendous implications for the nature of DE.~\footnote{See for instance Refs.~\cite{Tada:2024znt,Yin:2024hba,Wang:2024hks,Luongo:2024fww,Cortes:2024lgw,Clifton:2024mdy,Colgain:2024xqj,Carloni:2024zpl,Wang:2024rjd,Allali:2024cji,Giare:2024smz,Wang:2024dka,Gomez-Valent:2024tdb,Yang:2024kdo,Park:2024jns,Escamilla-Rivera:2024sae,Wang:2024pui,Shlivko:2024llw,Dinda:2024kjf,Bousis:2024rnb,Seto:2024cgo,Favale:2024sdq,Croker:2024jfg,Bhattacharya:2024hep,Ramadan:2024kmn,Pogosian:2024ykm,Roy:2024kni,Jia:2024wix,Wang:2024hwd,Heckman:2024apk,Gialamas:2024lyw,Notari:2024rti,Lynch:2024hzh,Liu:2024txl,Chudaykin:2024gol,Dwivedi:2024okk,Liu:2024gfy,Orchard:2024bve,Patel:2024odo,Hernandez-Almada:2024ost,Pourojaghi:2024tmw,Mukhopadhayay:2024zam,Elbers:2024sha,Wang:2024sgo,Naredo-Tuero:2024sgf,Li:2024qso,Ye:2024ywg,Giare:2024gpk,Dinda:2024ktd,Jiang:2024viw} for examples of recent studies on implications of the DESI results for fundamental physics.}

When it comes to gathering clues on the cause of cosmic acceleration, a key issue to address is whether or not the DE energy density $\rho_x(z)$ is consistent with being constant in time (given that any reliable deviation from this picture would indicate that cosmic acceleration cannot be ascribed to a CC, and different strategies can be adopted to address this problem. Examples include model-dependent approaches, i.e.\ constraining specific models of DE or modifications to General Relativity, or parametric approaches, which posit (typically) simple parametric forms for $w_x(z)$ and/or $\rho_x(z)$, whose parameters are then confronted against observations: a difficulty with these approaches, however, is to come up with models and parametrizations which are well-motivated, or at least not inherently ad hoc. A more data-driven approach is to construct an ``optimal'' local basis representation for $\rho_x(z)$ using principal component analysis. Another approach, which has gained significant traction over the past decade, is to use a distribution over (random) functions to smooth $\rho_x(z)$ or $w_x(z)$, for instance through so-called Gaussian Processes (GPs), and estimate the statistical properties thereof in light of observed data. Such a nonparametric reconstruction approach carries several distinct advantages: for instance, it may recover behaviours which are unexpected a priori and cannot therefore be covered by parametric approaches, and allows one to be as model-independent (or rather nonparametric) as possible, while looking for deviations from the CC regardless of origin~\cite{Holsclaw:2010sk,Shafieloo:2012ht,Seikel:2012uu}.

Our goal in this work fits precisely in the above picture. In particular, using the latest DESI BAO data, we aim to reconstruct the $z \lesssim 2$ expansion history, while also examining the implications of our results for the Hubble tension from our nonparametric inference of $r_dH_0$ (with $r_d$ the sound horizon at baryon drag and $H_0$ the Hubble constant). To carry out the reconstruction, we adopt two approaches: a cubic spline interpolation following the approach of Refs.~\cite{Bernal:2016gxb,Aylor:2018drw}, and a GP-based reconstruction. This allows us to provide model-independent insight into the DESI hints for evolving DE (which we cross-check against earlier BAO measurements with a similar analysis procedure), as well as the role of external SNeIa datasets, while confirming the impossibility of solving the Hubble tension with late-time smooth modifications to the expansion history~\cite{Bernal:2016gxb,Addison:2017fdm,Lemos:2018smw,Aylor:2018drw,Schoneberg:2019wmt,Knox:2019rjx,Arendse:2019hev,Efstathiou:2021ocp,Cai:2021weh,Keeley:2022ojz}. In particular, our results nonparametrically identify the presence of two features in the unnormalized expansion rate $E(z)$ potentially ascribable to deviations from the cosmological constant picture: a bump at $z \sim 0.5$, and a depression at $z \sim 0.9$.

The rest of this paper is then organized as follows. We briefly introduce our methodology and adopted datasets in Sec.~\ref{sec:methodologydatasets}. The results of our analysis and discussion thereof are presented in Sec.~\ref{sec:results}. We then draw concluding remarks in Sec.~\ref{sec:conclusions}. A brief discussion on the impact of changing the interpolation knots in the cubic spline reconstruction procedure is presented in Appendix~\ref{appendix:knots}, whereas a similar discussion on the impact of the GP kernel is carried out in Appendix~\ref{appendix:kernel}.

\section{Methodology and datasets}
\label{sec:methodologydatasets}

\subsection{Methodology}
\label{subsec:methodology}

We now discuss the methodology adopted to simultaneously nonparametrically reconstruct/constrain both the late-time ($z \lesssim 2$), unnormalized expansion history of the Universe and the combination $r_dH_0$. As anticipated previously, we will do so by making use of two different methods: a cubic spline interpolation, following Refs.~\cite{Bernal:2016gxb,Aylor:2018drw}, and a Gaussian Process-based Markov Chain Monte Carlo (MCMC) analysis, following the method introduced in Ref.~\cite{Ye:2023zel}.

\subsubsection{Cosmological background}
\label{subsubsec:cosmological}

We begin by briefly introducing the relevant cosmological quantities we will use in the subsequent analysis. Our goal is for our conclusions to be as model-independent as possible. To this end, we build upon minimal cosmological assumptions: a homogeneous and isotropic cosmology described by a (not necessarily spatially flat) Friedmann-Lema\^{i}tre-Robertson-Walker (FLRW) space-time, and the Etherington distance-duality relation (DDR)~\cite{Etherington:1933ghw}.~\footnote{We note, however, that the Hubble tension may ultimately require departures from the FLRW framework~\cite{Krishnan:2021dyb}.} In addition, we are implicitly assuming that the late-time expansion history is sufficiently smooth.

Transverse comoving distances (also referred to as comoving angular diameter distances), which we denote by $D_M$, are given by the following:
\begin{eqnarray}
D_M(z) = \frac{c}{H_0\sqrt{\vert \Omega_K \vert}} \text{sinn} \left ( \sqrt{\vert \Omega_K \vert} \int_0^z\, \frac{dz'}{E(z')} \right ) \,,
\label{eq:dm}
\end{eqnarray}
where $H_0$ is the Hubble constant, $\Omega_K$ corresponds to the present-day spatial curvature parameter (with $\Omega_K<0$ and $\Omega_K>0$ corresponding to a spatially closed and spatially open Universe respectively), $E(z) \equiv H(z)/H_0$ denotes the unnormalized expansion rate (one of the quantities we aim to nonparametrically reconstruct), and the function ${\text{sinn}}(x)$ is given by ${\text{sinn}}(x)=\{\sin(x), \sinh(x),x\}$ for $\{\Omega_K<0,\Omega_K>0,\Omega_K=0\}$ respectively.

A key quantity relevant for BAO measurements is the sound horizon at baryon drag $r_d$, which leaves its imprint in the form of a localized peak in the two-point correlation function of tracers of the LSS, and is given by the following expression:
\begin{eqnarray}
r_d = \int_{z_d}^{\infty}dz\, \frac{c_s(z)}{H(z)}\,,
\label{eq:rd}
\end{eqnarray}
where $z_d \approx 1060$ is drag epoch, when baryons were released from the drag of photons. In what follows, we will assume that $z_d$ remains fixed to this value. Transverse BAO measurements extracted from a LSS tracer sample located at an effective redshift $z_{\text{eff}}$ are sensitive to the BAO angular scale $\theta_d$, which sets a preferred angular separation scale between pairs of galaxies (for separation vectors located transverse to the observer's line-of-sight) and is given by the following:
\begin{eqnarray}
\theta_d(z_{\text{eff}}) = \frac{r_d}{D_M(z_{\text{eff}})} = \frac{r_dH_0}{c \int_0^{z_{\text{eff}}} dz'/E(z')}\,.
\label{eq:thetad}
\end{eqnarray}
For separation vectors parallel to the observer's line-of-sight, and thereby line-of-sight BAO measurements, a preferred separation redshift $\delta z_d$ is instead observed, and is related to the Hubble distance $D_H(z)$ as follows:
\begin{eqnarray}
\delta z_d(z_{\text{eff}}) = \frac{r_d}{D_H(z_{\text{eff}})} = r_dH(z_{\text{eff}})/c=r_dH_0E(z_{\text{eff}})/c\,.
\label{eq:deltazd}
\end{eqnarray}
Finally, isotropic BAO measurements (in the case of samples whose signal-to-noise ratio is not sufficient to disentangle line-of-sight and transverse modes at high significance) are sensitive to the volume-averaged angular scale $\theta_v$, given by the following:
\begin{eqnarray}
\theta_v(z_{\text{eff}}) = \frac{r_d}{D_V(z_{\text{eff}})} = \frac{r_d}{ \left [ z_{\text{eff}} D_M^2(z_{\text{eff}})D_H(z_{\text{eff}}) \right ] ^{1/3}}\,.
\label{eq:thetav}
\end{eqnarray}
When measured across a sufficiently large effective redshift range, BAO measurements, typically reported in the form of $D_M/r_d$, $D_H/r_d$, and $D_V/r_d$ (or the reciprocals thereof), are therefore on their own (i.e.\ in the absence of an external calibration of $r_d$) sensitive to the unnormalized expansion history $E(z)$. In the minimal $\Lambda$CDM model, this translates into a particular sensitivity to $\Omega_m$. If appropriately calibrated, for instance through a Big Bang Nucleosynthesis (BBN) prior on the physical baryon density $\omega_b$, which in turn can be used to determine the sound horizon $r_d$, BAO instead become sensitive to the overall expansion rate $H(z)$, and therefore to $H_0$. We note that, within the minimal set of assumptions adopted, BAO measurements are characterized by three quantities: an overall (dimensionless) amplitude $\beta_{\text{BAO}} \equiv c/r_dH_0$, the unnormalized expansion rate $E(z)$, and the present-day spatial curvature parameter $\Omega_K$.

The unnormalized expansion rate can also be probed via SNeIa. In the absence of knowledge about their absolute magnitude, the evolution of the apparent magnitude of SNeIa at different redshifts can be used as a relative distance indicator to constrain $E(z)$. More specifically, SNeIa distance moduli measurements $\mu$ are related to the luminosity distance $D_L$ as follows:
\begin{eqnarray}
\mu(z) = m_B-M_B = 5\log_{10} \left [ \frac{D_L(z)}{{\text{Mpc}}} \right ] +25\,,
\label{eq:muz}
\end{eqnarray}
where $m_B$ is the observed SNeIa light-curve B-band rest-frame peak magnitude, and $M_B$ is a nuisance parameter characterizing the SNeIa absolute magnitude in the same band. It is clear from Eq.~(\ref{eq:muz}) that, in the absence of a calibration for $M_B$ (which can be achieved for instance through Cepheids or the Tip of the Red-Giant branch among others~\cite{Freedman:2017yms,Riess:2019qba}), high-$z$ SNeIa are sensitive to the uncalibrated luminosity distance $H_0D_L(z)$. Under our assumption that the Etherington DDR holds (which, we recall, is the only other assumption alongside that of an FLRW metric, making our subsequent results highly model-independent), this is given by the following:
\begin{eqnarray}
H_0D_L(z) &=& H_0(1+z)D_M(z) \nonumber \\
&=& \frac{c(1+z)}{\sqrt{\vert \Omega_K \vert}} \text{sinn} \left ( \sqrt{\vert \Omega_K \vert} \int_0^z\, \frac{dz'}{E(z')} \right ) \,,
\label{eq:dl}
\end{eqnarray}
which shows that uncalibrated SNeIa are sensitive to the unnormalized expansion rate $E(z)$ and the spatial curvature parameter $\Omega_K$. Two comments are necessary before moving on. First, in the earlier discussion, for simplicity we neglected the impact of the usual stretch and color corrections. These would enter the second equality in Eq.~(\ref{eq:muz}), but do not impact the overall conclusions, and are taken care of in the subsequent analysis. Finally, we note that the Etherington DDR follows from the metricity of the underlying theory of gravity (i.e.\ the assumption that the Christoffel symbols can be expressed via the metric tensor, which implies that the manifold is pseudo-Riemannian), as well as the masslessness and number conservation of photons. While these appear to be fairly general assumptions and the Etherington DDR is in good agreement with current data~\cite{Santos-da-Costa:2015kmv,Hogg:2020ktc,Renzi:2021xii,Liu:2023hdz,Qi:2024acx}, we note that the duality can be broken in theories where photons are not massless (and therefore do not travel along null geodesics) or where photon number is not conserved (for instance in the presence of decays to other particles), or in theories whose space-time is not a pseudo-Riemannian manifold, such as $f(Q)$ theories~\cite{BeltranJimenez:2019esp,Heisenberg:2023lru}: our results do not (necessarily) apply to these frameworks.

In what follows, we combine BAO and uncalibrated SNeIa measurements to nonparametrically reconstruct the late-time unnormalized expansion history of the Universe, $E(z \lesssim 2)$, in both spatially flat and nonflat Universes. The dataset combination in question helps to break the degeneracy between $r_dH_0$ and $E(z)$, whereas the degeneracy between $r_d$ and $H_0$ can only be further broken by inputting additional information on either $r_d$ (from the early Universe) or $H_0$ (from the local Universe). We now turn to discuss the methods adopted to perform the reconstruction.

\subsubsection{Cubic spline interpolation}
\label{subsubsec:cubicspline}

Inspired by the seminal works of Bernal, Verde, and Riess~\cite{Bernal:2016gxb}, and Aylor \textit{et al.}~\cite{Aylor:2018drw}, the first method we adopt is a cubic spline interpolation. More specifically, $E(z)$ is expressed as a combination of piece-wise cubic splines. The reconstructed $E(z)$ is specified by the values it takes at five redshift knots, $z_i=\{0.2, 0.57, 0.8, 1.3, 2.33\}$ (with an additional knot at present time such that $E(z=0)=1$ following the definition of $E$), and at these knots the first and second derivatives of the splines are continuous, leading to a twice continuously differentiable curve. With the boundary condition that the second derivatives vanish at two boundary knots (therefore these are technically natural cubic splines), the piece-wise natural cubic spline is mathematically uniquely defined.

For what concerns the choice of knots, the first four are the same ones adopted by Refs.~\cite{Bernal:2016gxb,Aylor:2018drw}. On the other hand, the $z=2.33$ one has been added to account for new BAO and SNeIa data at higher redshifts which have become available since these earlier works. Nevertheless, as explicitly shown in Ref.~\cite{Aylor:2018drw}, the exact position of the knots has a negligible impact on the final results, and we confirm this conclusion in Appendix~\ref{appendix:knots}. More details on the exact MCMC implementation of the interpolation are given in Sec.~\ref{subsec:data}.

\subsubsection{Gaussian Process-based reconstruction}
\label{subsubsec:gpbased}

The second method we adopt is a Gaussian Process-based reconstruction, very similar in spirit to that adopted in Ref.~\cite{Ye:2023zel} to nonparametrically reconstruct the shape of the CMB lensing potential. We recall that GPs generalize the concept of Gaussian distributions to the (infinite-dimensional) function space, in such a way that any finite collection of points of a given function $f(x)$ follows a multivariate Gaussian distribution controlled by the mean $\mu(x)$ and covariance function $k(x,\tilde{x})$~\cite{Rasmussen:2006ghw}. We indicate this as follows:
\begin{eqnarray}
f(x) \sim {\cal GP} \left ( \mu(x),k(x,\tilde{x}) \right ) \,.
\label{eq:gp}
\end{eqnarray}
Over the past decade, after their first seminal applications~\cite{Holsclaw:2010sk,Shafieloo:2012ht,Seikel:2012uu}, GPs have found widespread use in cosmology, in the context of nonparametric reconstructions of relevant cosmological functions (typically the redshift evolution of quantities such as distances, expansion rates, densities, equations of state, growth rates, and so on), see e.g.\ Refs.~\cite{Zhang:2018gjb,Elizalde:2018dvw,Gerardi:2019obr,LHuillier:2019imn,Cai:2019bdh,Liao:2019qoc,Aljaf:2020eqh,Benisty:2020kdt,Briffa:2020qli,Keeley:2020aym,Renzi:2020bvl,Bonilla:2020wbn,Renzi:2020fnx,vonMarttens:2020apn,OColgain:2021pyh,Bonilla:2021dql,Bora:2021bui,Bora:2021cjl,Dhawan:2021mel,Mukherjee:2021kcu,Sun:2021pbu,Escamilla-Rivera:2021rbe,Bernardo:2021qhu,Bernardo:2021mfs,Bengaly:2021wgc,Jesus:2021bxq,Ruiz-Zapatero:2022zpx,Avila:2022xad,Wang:2022rvf,Benisty:2022psx,Mukherjee:2022ujw,Ren:2022aeo,Perenon:2022fgw,Hwang:2022hla,Wu:2022fmr,Li:2022cbk,Calderon:2023msm,Calderon:2023obf,Mukherjee:2023lqr,Dinda:2023svr,Escamilla:2023shf,Dinda:2023xqx,Favale:2024lgp,Barbosa:2024ppn,Dinda:2024xla,Yang:2024tkw,Zhang:2024ndc,Mukherjee:2024ryz,Sabogal:2024qxs,Mukherjee:2024cfq}.

The function $k(x,\tilde{x})$ is typically referred to as kernel, and controls the strength of correlations between the values of the reconstructed function at different points, as well as the deviations from the mean at any given point. The range of possible covariance functions is extremely wide, with most of them depending on the distance between input points $\vert x-\tilde{x} \vert$ (though, we stress, this is not a necessary requirement). In what follows, we adopt the widely used squared exponential kernel (also known as radial basis function), defined as follows:
\begin{eqnarray}
k(x_i, x_j) = \sigma^2 \exp \left ( - \frac{(x_i - x_j)^2}{2\ell^2} \right ) \,.
\label{eq:squaredexponentialkernel}
\end{eqnarray}
The squared exponential kernel is characterized by two hyperparameters: the correlation length $\ell$, which characterizes the overall ``smoothness''/``wiggliness'' of the reconstructed function, and the output variance $\sigma^2$, which controls the average distance of the reconstructed function away from its mean. This kernel represents a widely used, standard choice in applications of GP reconstruction in cosmology. There are various reasons behind this choice, which are also relevant for our case. Firstly, not only is the kernel infinitely differentiable (making each realization of the associated GP infinitely differentiable), but it is also characterized by infinite support. In addition, the kernel enjoys various useful mathematical properties, including its being universal (i.e.\ it can be used to approximate an arbitrary continuous target function uniformly on any compact subset of the input space) and integrable against any function. Finally, its depending on only two hyperparameters makes it highly manageable. Overall, the squared exponential kernel is therefore highly suited for capturing global features of smooth, continuous functions. The reasonable and commonly held belief that functions of cosmological interest such as distances and expansion rates should be smooth at late times therefore explains the success enjoyed by this kernel. We shall also adopt the squared exponential kernel in our work with the same motivations discussed above, noting that this implicitly entails our setting smoothness requirements on the functions we attempt to reconstruct.~\footnote{For this reason, our choice of kernel may not be appropriate for models where the expansion rate is inherently discontinuous. An example in this sense which has gained significant interest recently is the so-called $\Lambda_s$CDM model, featuring an instantaneous sign-switch in the cosmological constant~\cite{Akarsu:2019hmw,Acquaviva:2021jov,Akarsu:2021fol,Akarsu:2022typ,Akarsu:2023mfb,Paraskevas:2024ytz,Akarsu:2024qsi,Akarsu:2024eoo,Yadav:2024duq,Toda:2024ncp}.} Nevertheless, in Appendix~\ref{appendix:kernel} we study the impact of other choices of kernels, specifically the Mat\'{e}rn-$7/2$ and Mat\'{e}rn-$9/2$ kernels, finding that they lead to essentially the same results.

The standard usage of GPs in cosmology, which can be referred to as ``GP regression'', attempts to directly reconstruct functions of cosmological interest by adding information from observed data to obtain a conditional GP distribution. However, strictly speaking this requires the data to also be described by a GP, and therefore for the errors to be Gaussian \textit{in the variable one is trying to reconstruct} (see e.g.\ the discussion surrounding Eq.~(2.8) of Ref.~\cite{Seikel:2012uu}). This can be problematic if, for instance, one is trying to reconstruct the expansion rate, since most of the BAO measurements are Gaussian in distances and therefore not in the expansion rate, given that the relation between the two is a nonlinear (integral) one. This issue has gone unappreciated in part of the cosmological literature, while other works take it into account by working only with a subset of available data -- for instance, if one is interested in reconstructing $H(z)$, only line-of-sight BAO data would be used among the many available BAO measurements. However, this implies not harnessing the full constraining power of available cosmological data. To overcome this problem, we adopt a different strategy (which for simplicity we refer to as GP-based reconstruction), first used by one of us in Ref.~\cite{Ye:2023zel}. Specifically, we start from a reference unnormalized expansion history $E_{\text{ref}}(z)$, which is then related to the $E(z)$ we wish to reconstruct as follows:
\begin{eqnarray}
E(z) = e^{A(z)} E_{\text{ref}}(z),
\label{eq:egp}
\end{eqnarray}
where it is the function $A(z)$ itself, characterizing the (logarithmic) deviations of $E(z)$ from the reference model, which follows a GP with zero mean and squared exponential kernel:
\begin{eqnarray}
A(z) \sim {\cal GP} \left ( 0,\sigma^2 \exp \left [ - \frac{(z - \tilde{z})^2}{2\ell^2} \right ] \right ) \,.
\label{eq:a}
\end{eqnarray}
In this way, the GP mean corresponds to $E(z)=E_{\text{ref}}(z)$, whereas the limits $A(z) \to -\infty$ and $A(z) \to \infty$ correspond to $E(z) \to 0$ and $E(z) \to \infty$ respectively.~\footnote{We adopt the parametrization in Eq.~(\ref{eq:egp}) since, a priori, we do not know the order of magnitude of the deviations from the reference model: this is somewhat analogous to cases where one may prefer to set a uniform prior on the logarithm of a parameter rather than on the parameter itself. Nevertheless, a posteriori the deviations from the reference model are found to not exceed $\sim 10\%$ in $E(z)$. In this case, one may expand the exponential to first order, with the sum of the resulting two terms playing the role of overall coefficient $A(L)$ in Eq.~(1) of Ref.~\cite{Ye:2023zel}, and the first order term following a GP distribution with zero mean.} To aid the sampling process, we follow Ref.~\cite{Ye:2023zel} and introduce six nodes, evenly distributed in $z \in [0, 2.33]$, while setting the boundary condition $A(z=0)=0$ so that $E(z=0)=E_{\text{ref}}(z=0)=1$.

Before moving on to discuss the adopted datasets and analysis methodology, a general comment on GPs is in order to better qualify the outcome of our work. In the literature, it is common practice to refer to GPs as being model-independent and nonparametric, neither of which is, strictly speaking, correct. The results are certainly independent of any explicit cosmological model, but may depend on the assumed kernel and its properties (see e.g.\ Ref.~\cite{Zhang:2023pis} for a recent discussion on kernel selection for GPs in cosmology). As discussed earlier we check the impact of other widely used kernels in Appendix~\ref{appendix:kernel}, finding it to be negligible, but it is strictly speaking impossible to demonstrate that the results are stable against \textit{any} possible kernel. In addition, GPs are to some extent parametric or, more precisely, hyperparametric. Nevertheless, this is somewhat tamed by our choice of marginalizing over the hyperparameters (more details shortly in Sec.~\ref{subsec:data}). In short, while GPs are without any doubt less model-dependent and parametric compared to other approaches in cosmology, their model-dependence and (hyper)parametric nature is somewhat more latent but still present (see also Ref.~\cite{Gomez-Valent:2018hwc} for more details on these points). With these caveats in mind, in what follows we will generally refrain from referring to our results as being model-independent, but will still refer to them as being nonparametric, with what we feel is only a (very) slight abuse of terminology.

\subsection{Datasets}
\label{subsec:data}

To carry out the cubic spline and GP-based reconstruction procedures, we make use of various combinations of BAO and (uncalibrated) SNeIa datasets. As discussed in Sec.~\ref{subsubsec:cosmological}, these combinations are sufficient to disentangle the effects of the dimensionless amplitude $\beta_{\text{BAO}}=c/r_dH_0$, the unnormalized expansion rate $E(z)$, and the spatial curvature parameter $\Omega_K$. At a later stage, we include a prior on the local value of $H_0$ to further break the degeneracy between $r_d$ and $H_0$.

For what concerns BAO measurements, we make use of the following datasets.
\begin{itemize}
\item \textbf{DESI} -- We adopt the latest BAO measurements from Data Release 1 of the Dark Energy Spectroscopic Instrument (DESI)~\cite{DESI:2024mwx}, obtained from various samples of LSS tracers: the Bright Galaxy Sample (BGS) at an effective redshift $z_{\text{eff}}=0.295$; three Luminous Red Galaxy (LRG) samples (LRG1, LRG2, and LRG3) at $z_{\text{eff}}=0.510$, $0.706$, and $0.930$; two Emission Line Galaxy (ELG) samples (ELG1 and ELG2) at $z_{\text{eff}}=0.930$, $1.317$; the Quasar (QSO) sample at $z_{\text{eff}}=1.491$; and the Lyman-$\alpha$ Forest (Ly-$\alpha$) sample at $z_{\text{eff}}=2.330$. Following the DESI analysis~\cite{DESI:2024mwx}, we do not consider the LRG3 and ELG1 samples separately, but the combined LRG3+ELG1 sample. We refer the reader to Ref.~\cite{DESI:2024mwx} for further details on these samples and the associated BAO measurements. 
\item \textbf{SDSS} -- For comparison, we also adopt earlier BAO measurements from the Baryon Oscillation Spectroscopic Survey (BOSS) and extended BOSS (eBOSS) survey programs of the Sloan Digital Sky Survey (SDSS), and in particular from: the Main Galaxy Sample (MGS) at $z_{\text{eff}}=0.15$; the BOSS galaxy samples at $z_{\text{eff}}=0.38$, $0.51$; the eBOSS LRG sample at $z_{\text{eff}}=0.70$; the eBOSS ELG sample at $z_{\text{eff}}=0.85$; the eBOSS QSO sample at $z_{\text{eff}}=1.48$; the eBOSS Ly-$\alpha$ sample and the cross-correlation between the Ly-$\alpha$ and QSO samples, both at $z_{\text{eff}}=2.33$. We note that, in order for the comparison against DESI data to be as fair as possible, we only use distance and expansion rate measurements -- $D_V/r_d$, $D_M/r_d$, and $D_H/r_d$ and associated quantities -- and not growth rate ($f\sigma_8$) measurements. We refer the reader to Ref.~\cite{eBOSS:2020yzd} for further details on these BAO measurements.
\end{itemize}
In what follows, we make use of the DESI and SDSS datasets separately, without combining them.

The adopted BAO measurements are always combined with one of the following three SNeIa samples:
\begin{itemize}
\item \textbf{PantheonPlus} -- We adopt the \textit{Pantheon+} sample, consisting of 1701 light curves for 1550 unique SNeIa. We only use points in the redshift range $0.01<z<2.26$, which therefore does not include the calibration sample at lower redshifts~\cite{Scolnic:2021amr}.
\item \textbf{Union3} -- We also consider the Union3 sample obtained from a compilation of 2087 SNeIa from 24 datasets in the redshift range $0.01<z<2.26$, analyzed through an unified Bayesian framework and binned in $0.05<z<2.26$~\cite{Rubin:2023ovl} (note that we adopt binned distance moduli, as these are the only ones available for the moment).
\item \textbf{DESY5} -- Finally, we make use of the photometrically classified sample obtained during the full five years of the Dark Energy Survey (DES) Supernova Program, which contains 1635 SNeIa in the redshift range $0.1<z<1.3$, complemented by 194 high-quality external SNeIa from the CfA3~\cite{Hicken:2009df}, CfA4~\cite{Hicken:2012zr}, CSP~\cite{Krisciunas:2017yoe}, and Foundation~\cite{Foley:2017zdq} samples in the redshift range $0.025<z<0.1$~\cite{DES:2024tys}.
\end{itemize}
As with the BAO measurements, we only consider one SNeIa sample at a given time, without ever combining them (we also note that some SNeIa are common between the samples).

Finally, at the very last stage of our work, we aim to draw conclusions on the sound horizon and the Hubble tension, and to do so we need to break the degeneracy between $r_d$ and $H_0$, which enter multiplicatively into $\beta_{\text{BAO}}$. When the PantheonPlus sample is included, we achieve this by including the full calibrator SNeIa sample in the redshift range $0.023<z<0.15$, accounting for the covariance with the rest of the sample. When either the Union3 or DESY5 sample is included instead, we impose an external Gaussian prior on $H_0 = (73.04 \pm 1.04)\,\text{km/s/Mpc}$ as reported in Ref.~\cite{Riess:2021jrx}. We note that a more up-to-date SH0ES distance ladder measurement of $H_0 = (73.17 \pm 0.86)\,\text{km/s/Mpc}$ is available~\cite{Breuval:2024lsv}. However, switching to this prior would not significantly alter our conclusions.

\begin{figure*}[!ht]
\centering
\includegraphics[width=0.43\linewidth]{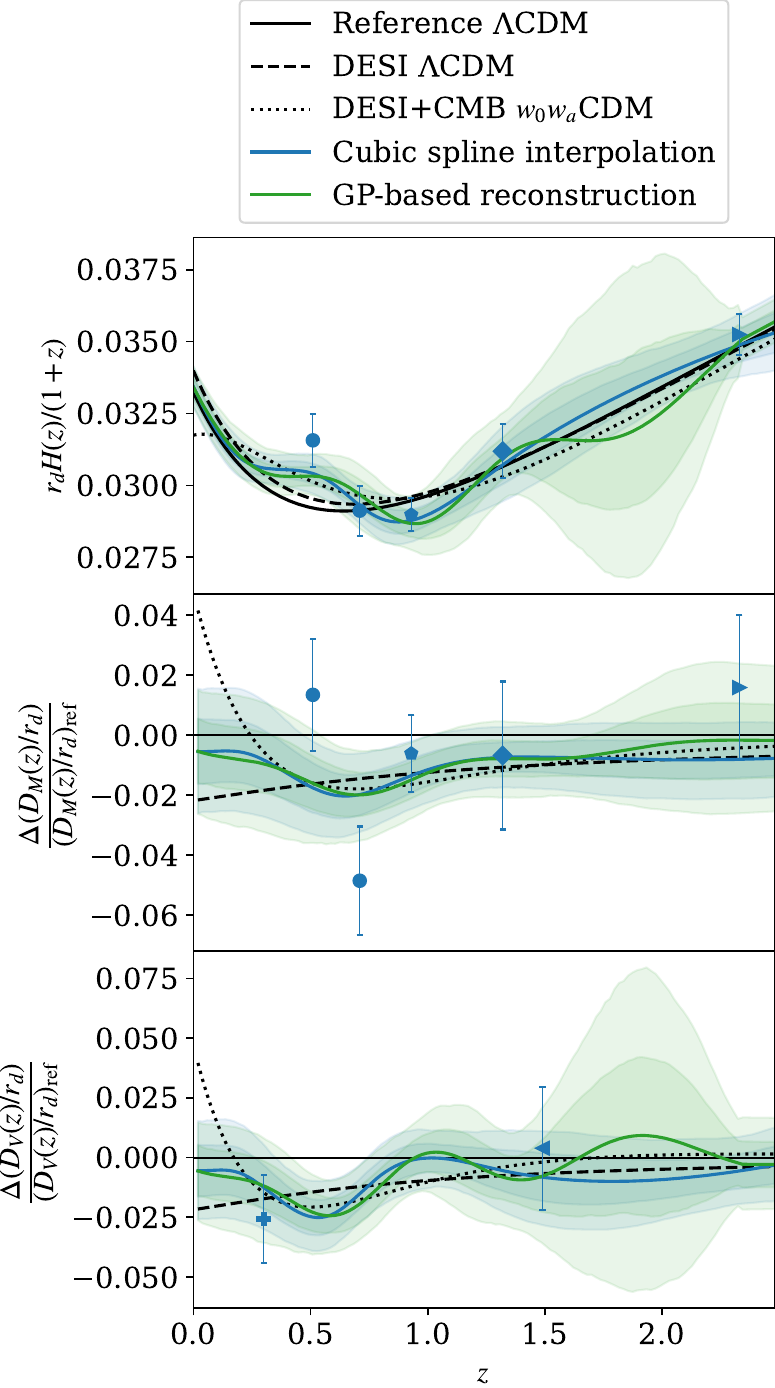}\,\,\,\,\,\,\,\,\,
\includegraphics[width=0.43\linewidth]{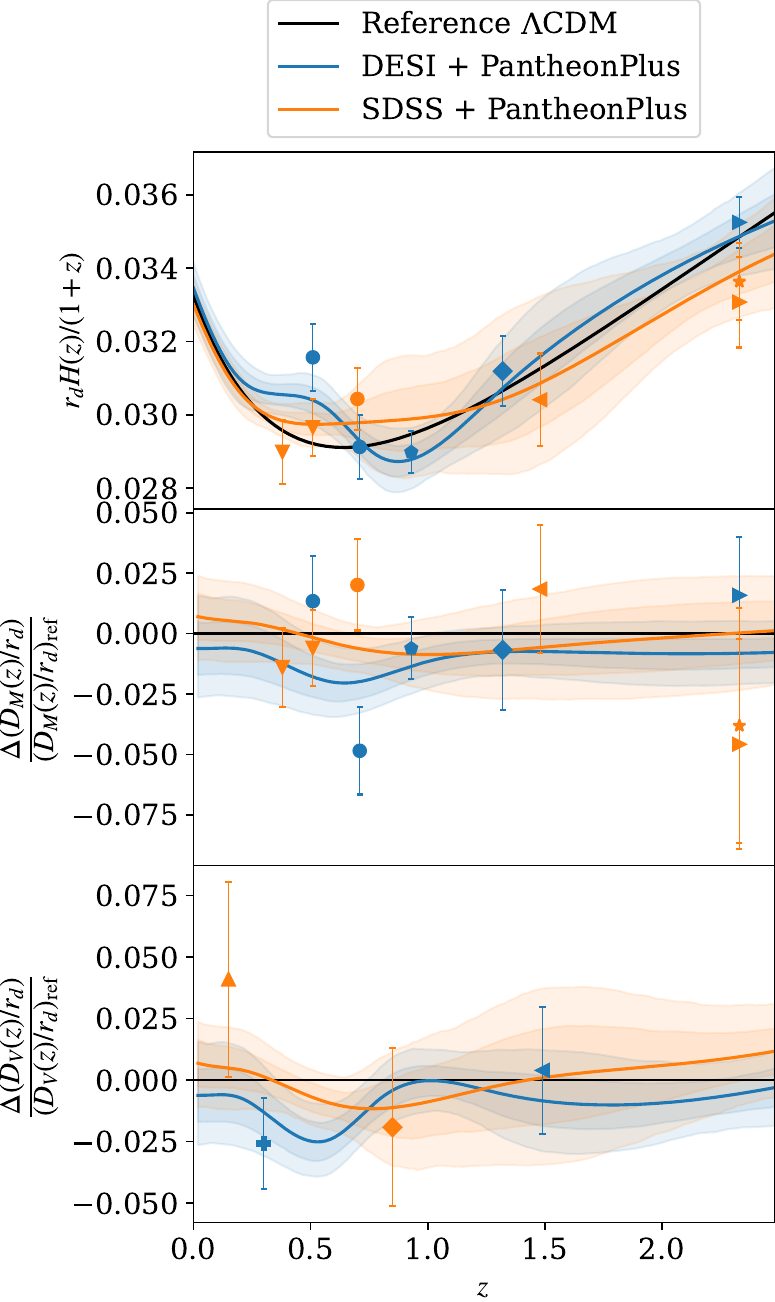}
\caption{\textit{Left panel}: reconstructed posteriors for $r_dH(z)/(1+z)$, $D_M(z)/r_d$, and $D_V(z)/r_d$. For the latter two we plot their relative deviation with respect to the reference $\Lambda$CDM model, namely $\Delta (D_M(z)/r_d)/(D_M(z)/r_d)_{\text{ref}}$ and $\Delta (D_V(z)/r_d)/(D_V(z)/r_d)_{\text{ref}}$. The blue curve has been obtained through cubic spline interpolation, whereas the green curve has been obtained via Gaussian Process-based reconstruction, in both cases using DESI+PantheonPlus data. In both cases, the dark and light bands indicate 68\% and 95\% credible intervals for the reconstructed functions, while the respective colored solid curves correspond to the reconstructed mean functions. Also plotted are the evolutions of the same functions for the reference $\Lambda$CDM model (solid black curve), for a $\Lambda$CDM model fit to DESI data (black dashed curve), and for a $w_0w_a$CDM model fit to DESI and CMB data (black dotted curve), alongside the appropriately rescaled DESI BAO measurements (blue datapoints). \textit{Right panel}: comparison between the results obtained using DESI (blue curves, bands, and datapoints) and SDSS (orange curves, bands, and datapoints) BAO measurements, with the same subfigure structure as the left panel. In both cases, PantheonPlus SNeIa data has been adopted, and cubic spline interpolation has been used.}
\label{fig:splinevsgpdesivssdss}
\end{figure*}

When performing the cubic spline interpolation, the free parameters we vary are the overall amplitude of the BAO scale(s) $\beta_{\text{BAO}}$, the values of the unnormalized expansion rate at the five knots $E_1$, $E_2$, $E_3$, $E_4$, and $E_5$ -- with $E_i = E(z_i)$ -- and finally the spatial curvature parameter $\Omega_K$, the latter only when we are not explicitly considering a spatially flat FLRW model. When we instead perform the GP-based reconstruction, the free parameters are again $\beta_{\text{BAO}}$, the values the GP-distributed function $A(z)$ appearing in Eq.~(\ref{eq:egp}) takes at the six knots $A_0$, $A_1$, $A_2$, $A_3$, $A_4$, and $A_5$ -- again with $A_i=A(z_i)$ and $A_0 = A(z=0) = 1$ -- and once more the spatial curvature parameter $\Omega_K$ when not explicitly considering a spatially flat FLRW model. As for the hyperparameters of the squared exponential kernel, we note that too large values of $\sigma$ and too small values of $\ell$ will worsen the convergence of the reconstruction, although $\sigma$ should still be larger than the typical uncertainties in the data. In what follows, we vary both $\sigma$ and $\ell$, although we note that following the widespread approach of pretraining them and fixing them to the trained values ($\ell \approx 0.3$ and $\sigma \approx 0.1$) would lead to largely identical results. 

Operationally, for the GP-based reconstruction, at each point of the MCMC (to be discussed shortly) we directly sample a different (GP-distributed) function $A(z)$, and therefore a different unnormalized expansion rate $E(z)$ through Eq.~(\ref{eq:egp}). The sampled expansion rate can then be appropriately integrated in order to be compared against BAO and SNeIa measurements as in Eqs.~(\ref{eq:thetad},\ref{eq:deltazd},\ref{eq:thetav},\ref{eq:dl}). Therefore, while our method remains overall nonparametric, at each step of the MCMC we still have a well-defined $E(z)$ which can be unambiguously compared against all the different types of datasets -- $D_M/r_d$, $D_H/r_d$, $D_V/r_d$, and $\mu(z)$.

To sample the posterior distributions of the cosmological parameters, we make use of MCMC methods. We adopt the cosmological MCMC sampler \texttt{Cobaya}~\cite{Torrado:2020dgo}, monitoring the convergence of the generated chains via the Gemlan-Rubin $R-1$ parameter~\cite{Gelman:1992zz} and requiring $R-1<0.01$ for our chains to be considered converged. The chains are analyzed via the \texttt{GetDist} package~\cite{Lewis:2019xzd}. The reference model adopted in our GP-based reconstruction, see Eq.~(\ref{eq:egp}), is chosen to be the $\Lambda$CDM model with best-fit parameters as reported in the \textit{Planck} 2018 cosmological parameters paper~\cite{Planck:2018vyg} and obtained from a fit to the \textit{Planck} 2018 TTTEEE, lowE, and lensing data, alongside then-current (BOSS DR12) BAO measurements. We note that in the GP-based reconstruction the resulting GP posterior turns out to be independent of the choice of location for the knots, as these only serve to help with the MCMC convergence (see Ref.~\cite{Ye:2023zel} for further details).

Our baseline dataset combination is the DESI+PantheonPlus one, whereas our baseline reconstruction method is the cubic spline interpolation. In all our subsequent analyses, we will show 68\% and 95\% credible intervals for $r_d H(z)/(1+z)$ as derived from the GP posterior of $A(z)$, and similarly for the relative variation in $D_M(z)/r_d$ and $D_V(z)/r_d$ with respect to the reference model reported above -- $\Delta (D_M(z)/r_d)/(D_M(z)/r_d)_{\text{ref}}$ and $\Delta (D_V(z)/r_d)/(D_V(z)/r_d)_{\text{ref}}$. In some cases, we will compare the resulting reconstructed functions against the best-fit predictions obtained from either $\Lambda$CDM or $w_0w_a$CDM fits to DESI and/or DESI+CMB data~\cite{DESI:2024mwx}.

\section{Results and discussion}
\label{sec:results}

\begin{figure*}[!ht]
\centering
\includegraphics[width=0.44\linewidth]{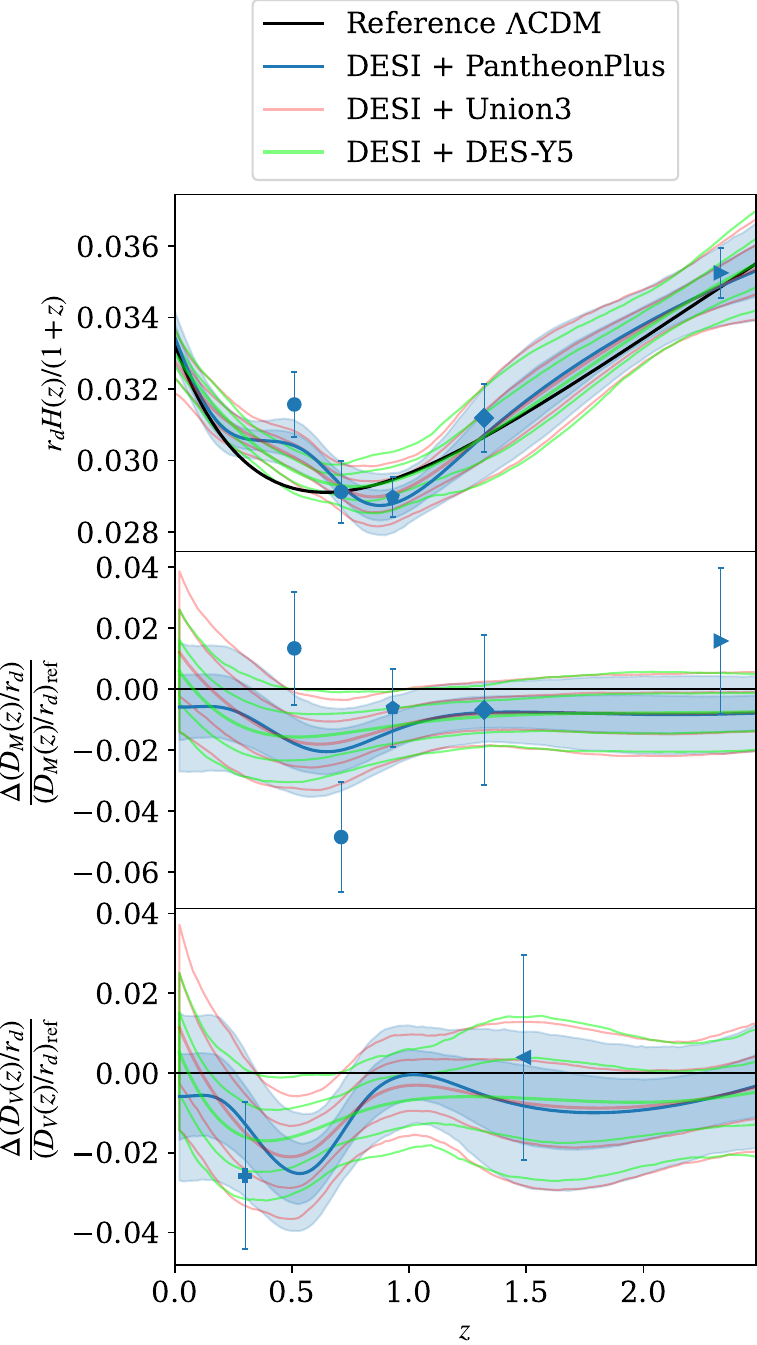}\,\,\,\,\,\,\,\,\,
\includegraphics[width=0.45\linewidth]{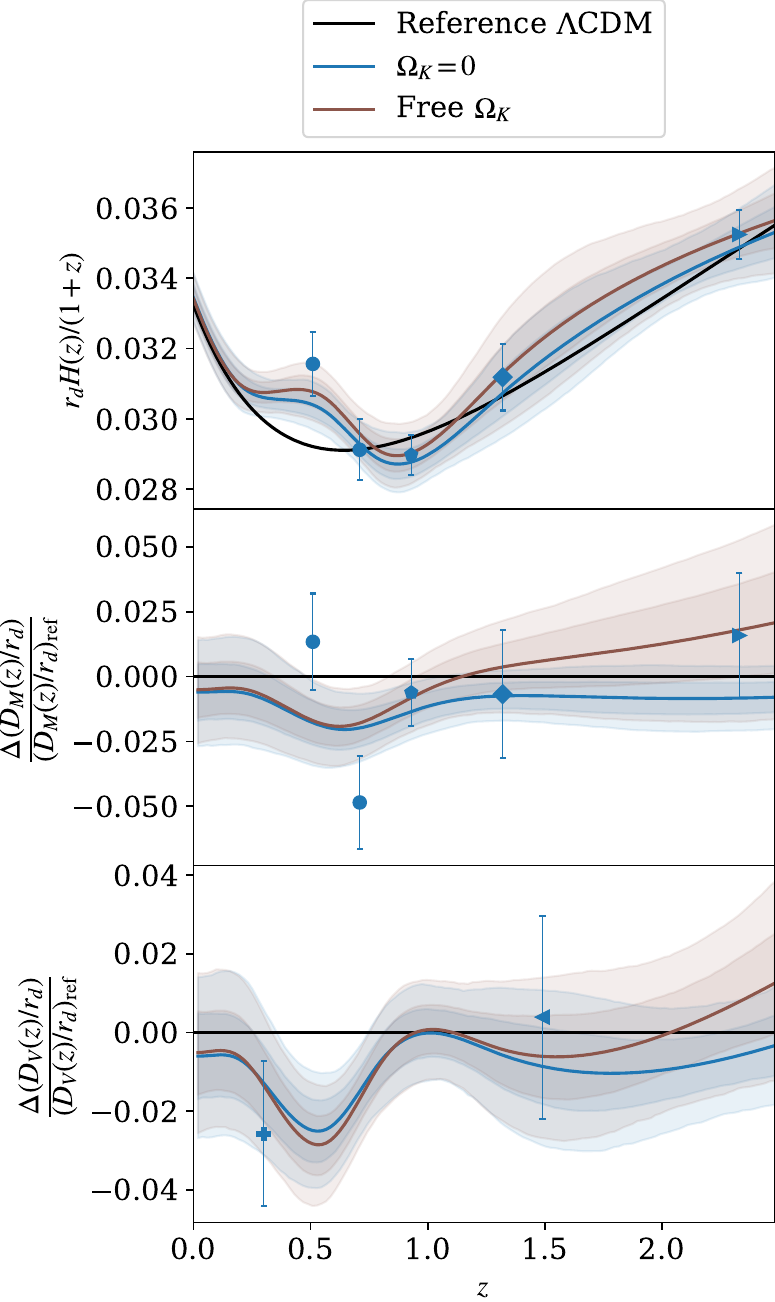}
\caption{\textit{Left panel}: comparison between the results obtained using the PantheonPlus (blue curves and bands), Union3 (pink curves), and DESY5 (green curves) SNeIa datasets, with the same subfigure structure as the left panel of Fig.~\ref{fig:splinevsgpdesivssdss}. In all cases, DESI BAO data has been adopted, and cubic spline interpolation has been used. \textit{Right panel}: comparison between the results obtained within a spatially flat FLRW model (blue curves and bands) and when allowing the spatial curvature parameter $\Omega_K$ to vary (brown curves and bands), with the same subfigure structure as the left panel. In both cases, the DESI+PantheonPlus dataset combination has been adopted, and cubic spline interpolation has been used.}
\label{fig:sneiacurvature}
\end{figure*}

We now discuss the results obtained using the methodology and datasets presented in Sec.~\ref{sec:methodologydatasets}. Visual summaries of our main results are provided in Figs.~\ref{fig:splinevsgpdesivssdss},~\ref{fig:sneiacurvature},~\ref{fig:hubbletension}, and~\ref{fig:ez}. Of these, the first two are composed of 3 subfigures, each showing the reconstructed redshift evolution of three background quantities: $r_dH(z)/(1+z)$ related to the (unnormalized) expansion rate (upper subfigures), and the (inverse) BAO angular scales $D_M(z)/r_d$ (intermediate subfigures) and $D_V(z)/r_d$ (lower subfigures), both relative to the same quantities in the reference model.

\subsection{Baseline reconstruction from DESI and PantheonPlus data}
\label{subsec:reconstructiondesipantheonplus}

We begin by considering the baseline reconstruction which only uses DESI BAO and PantheonPlus SNeIa measurements. The reconstructed redshift evolutions of $r_dH(z)/(1+z)$, $D_V(z)/r_d$, and $D_M(z)/r_d$ are shown in the left panel of Fig.~\ref{fig:splinevsgpdesivssdss} for both the cubic spline (blue curve and bands) and GP-based (green curve and bands) reconstruction methods. Overlain on the same plot are the redshift evolutions of the same quantities as predicted within the reference $\Lambda$CDM model, within a $\Lambda$CDM fit to DESI BAO data (black dashed curve), as well as a $w_0w_a$CDM fit to \textit{Planck} CMB and DESI BAO data (black dotted curve). Finally, each panel also shows the subset of the DESI BAO measurements which can be directly compared to the quantity whose redshift evolution is being plotted.

We immediately observe that the cubic spline and GP reconstructions on the whole agree with one another, broadly recovering the same features. The deviation between the two reconstructions is largest at $z \gtrsim 1.5$, while still remaining consistent well within $1\sigma$. This is in part due to the relatively large uncertainties for the GP reconstruction in that redshift range, itself a consequence of the lack of BAO points between the QSO and Ly-$\alpha$ measurements at $z_{\text{eff}}=1.491$ and $z_{\text{eff}}=2.330$ respectively.

Both the cubic spline- and GP-reconstructed expansion histories display important deviations with respect to the predictions of the reference $\Lambda$CDM model, as well as the $w_0w_a$CDM fit to \textit{Planck} and DESI data (the differences are obviously larger with respect to the former model). Two features in particular stand out. The first is a ``bump''/enhancement in the expansion rate at $z \sim 0.5$, whereas the second is a ``depression''/decrease in the expansion rate at $z \sim 0.9$, both of which can clearly be observed in the upper subfigure of the left panel of Fig.~\ref{fig:splinevsgpdesivssdss}. The former is associated to the LRG1 point at $z_{\text{eff}}=0.510$, and the latter to the LRG3+ELG1 point at $z_{\text{eff}}=0.930$. At the level of distances, since (at fixed $H_0$) larger $E(z)$ corresponds to lower distances and viceversa, these two features are respectively associated to lower and higher (inverse) BAO scales, as we can appreciate in the reconstructed BAO angular scales in the intermediate and lower panels on the left panel of Fig.~\ref{fig:splinevsgpdesivssdss}.

The bump at $z \sim 0.5$ deviates by more than $2\sigma$ with respect to the reference model and is the most prominent feature. It mainly results from the lower value of $D_H(z)/r_d$, and thereby a higher value of $r_dH(z)$, for the LRG1 point at $z \sim 0.5$, as well as a higher value of $D_M(z)/r_d$ for the LRG2 point at $z_{\text{eff}}=0.706$, as evident from the intermediate panel. We observe that this feature is not captured by a $\Lambda$CDM fit. When fitting DESI data to a $\Lambda$CDM model (black dashed curve), one finds a larger value of $\Omega_m$ relative that of the reference model (black solid curve). This increase in $\Omega_m$ goes precisely in the direction of increasing the expansion rate at low redshifts, and is accompanied by a slight increase in $r_dH_0$ to restore consistency at higher redshift (compare the black solid and dashed curves). However, neither of these two shifts are sufficient to fit the bump feature well within the $\Lambda$CDM model.

From the black dotted curve we can see that the bump feature is partially captured by the $w_0w_a$CDM fit to \textit{Planck}+DESI data. The reason is that a model with $w_0 \gtrsim -1$ and $w_a \lesssim -1$, as favored by DESI data, can produce a faster expansion at the onset of DE domination, without resulting in a faster expansion closer to the present time, therefore naturally leading to a bump feature. On the other hand, the depression feature at $z\sim 0.9$, mainly driven by the higher value of $D_H(z)r_d$ for the LRG3+ELG1 point at $z_{\text{eff}}=0.930$, is captured by neither the $\Lambda$CDM nor $w_0w_a$CDM models. The reason is that, once a fit to the statistically more significant bump feature is achieved (or at least attempted -- but failed -- through an increase in $\Omega_m$, as in the $\Lambda$CDM case), neither model possesses enough degrees of freedom/complexity to fit this other feature. In fact, as is clear from the left panel of Fig.~\ref{fig:splinevsgpdesivssdss}, fitting both features requires an oscillatory/nonmonotonic behaviour, which cannot be accommodated within $w_0w_a$CDM, and all the more so within $\Lambda$CDM.

\subsection{Impact of BAO dataset: DESI versus SDSS}
\label{subsec:impactbao}

The results discussed previously have been obtained utilizing DESI BAO and PantheonPlus SNeIa data. We now move on to examining the impact of adopting the SDSS BAO dataset in place of the DESI one, while maintaining PantheonPlus as our SNeIa dataset. To simplify the discussion, we stick to the cubic spline interpolation reconstruction given that, as observed earlier, both reconstruction methods are in agreement with each other and recover broadly similar features. The results of our analysis are shown in the right panel of Fig.~\ref{fig:splinevsgpdesivssdss}, with the same subfigure structure as in the left panel of the same Figure. The results adopting the DESI and SDSS datasets are given by the blue and orange bands respectively, and are accompanied by the subset of (DESI or SDSS as appropriate) BAO measurements which can be directly compared to the quantity whose redshift evolution is being plotted.

We see that neither of the two features identified previously, i.e.\ the bump at $z \sim 0.5$ and the depression at $z \sim 0.9$, are recovered when replacing the DESI BAO measurements with their SDSS counterparts. In fact, the reconstruction adopting the latter is perfectly consistent with the reference $\Lambda$CDM model within the entire redshift range, even in spite of the clear ``linear'' trend with the three DR12 galaxy data points, which has been argued in Ref.~\cite{Escamilla:2023oce} to be partially responsible for a weak preference for phantom DE when fitting \textit{Planck} CMB and SDSS BAO data. This leads us to the conclusion that the peculiar, oscillatory/nonmonotonic nature of the expansion history reconstructed from \textit{Planck}+DESI and discussed earlier, is (not unexpectedly) strongly driven by the DESI BAO data, and is absent when replacing the latter with the older SDSS BAO data.

The \textit{Planck}+DESI and \textit{Planck}+SDSS reconstructions differ the most (by $\approx 2\sigma$) at $z \lesssim 0.7$. These differences are driven by the significant mismatch between the respective measurements of $D_M/r_d$, with the measurements at $z \sim 0.7$ disagreeing at the level of $3\sigma$ or more.~\footnote{The significance of the tension for this measurement should be calculated accounting for the overlap in sky volume between DESI and SDSS. A very conservative estimate in this sense has been performed in Sec.~3.3 of Ref.~\cite{DESI:2024mwx}. Accounting for this correlation, the significance of the discrepancy in the $0.6 \lesssim z \lesssim 0.8$ redshift range has been gauged to be $\approx 3\sigma$.} This may simply be the result of a (not too unlikely) statistical fluctuation: indeed, despite the total number of spectra in the DESI sample exceeding $6$ million, the overall effective volume covered by DESI at redshifts $z \lesssim 0.8$ is still lower than that of the completed SDSS sample, partly due to the survey strategy meant to prioritize depth~\cite{DESI:2024mwx}. Nevertheless, upcoming DESI data releases will be able to clarify whether these discrepancies are physical or simply due to statistical fluctuations (a possibility which the latest version of Ref.~\cite{DESI:2024mwx} explicitly argues in favor of). At present, little more can be said about the origin of the disagreement between DESI and SDSS measurements.

\subsection{Impact of SNeIa dataset}
\label{subsec:impactsneia}

Recalling that all the results discussed so far were obtained utilizing the PantheonPlus catalog, we now examine the impact of the adopted SNeIa catalog, once more sticking to the cubic spline interpolation reconstruction procedure. We replace the PantheonPlus SNeIa catalog with the Union3 and DESY5 ones, while this time fixing the BAO measurements to the DESI ones. The results are shown in the left panel of Fig.~\ref{fig:sneiacurvature}, again with the same subfigure structure as in the left panel of Fig.~\ref{fig:splinevsgpdesivssdss}.

We find that using SNeIa catalogs other than PantheonPlus results in all the trends observed previously being washed out. In particular, traces of the earlier oscillatory/nonmonotonic behavior are barely visible in the DESI+Union3 reconstruction (red bands) and are totally absent in the  DESI+DESY5 reconstruction (purple bands). More specifically, the $z \sim 0.9$ depression is no longer present, whereas the significance of the $z \sim 0.5$ bump is significantly reduced, although the preference for an enhanced expansion rate [higher $r_dH(z)$] in the $0.4 \lesssim z \lesssim 0.6$ redshift range persists. The latter is clearly driven by the LRG1 (low) $D_H/r_d$ point at $z \sim 0.5$, as we already saw earlier when switching from the DESI BAO dataset to the SDSS one, and is therefore unrelated to the adopted SNeIa catalog.

The unnormalized expansion history reconstructed from DESI+Union3 still exhibits $\approx 2\sigma$ deviations from the reference $\Lambda$CDM model, especially around the $z \sim 0.5$ bump feature. On the other hand, the DESI+DESY5 reconstruction appears to be consistent within $2\sigma$ with the reference $\Lambda$CDM model across the whole redshift range. Part of the difference, and in particular the ``stiffness'' of the curves, may be related to the lower number of samples of the PantheonPlus catalog in the $0.4 \lesssim z \lesssim 0.5$ redshift range (see e.g.\ Fig.~3 of Ref.~\cite{DES:2024tys}), allowing for more freedom in the reconstructed expansion history, which is instead more tightly constrained by the Union3 and DESY5 SNeIa samples (see however Ref.~\cite{Colgain:2024ksa}). Overall, our conclusion is that the choice of SNeIa dataset is of moderate importance in the reconstruction.

\begin{figure*}[!t]
\centering
\includegraphics[width=0.9\linewidth]{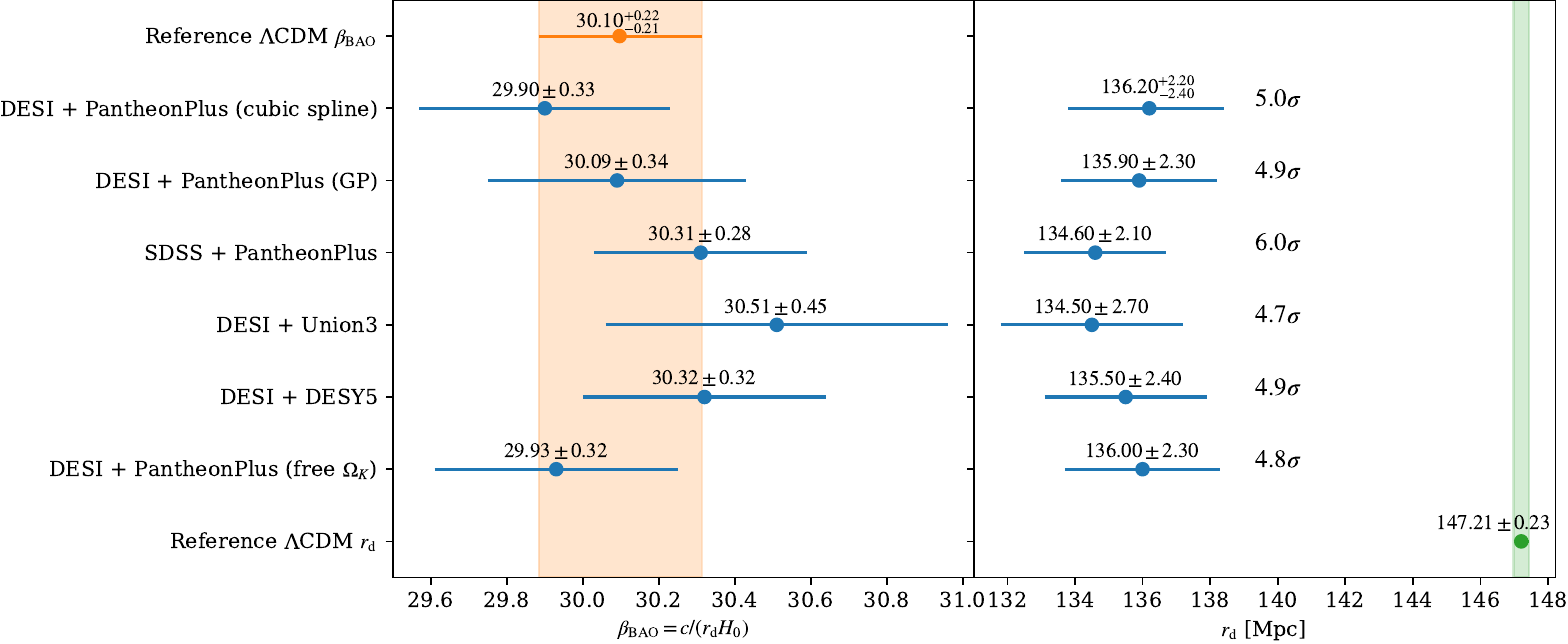}
\caption{Values of the dimensionless parameter $\beta_\text{BAO} \equiv c/r_dH_0$ (left side) and sound horizon at baryon drag $r_d$ (right side) inferred within the different analyses settings discussed in our work. If not explicitly specified, the results are obtained using cubic spline interpolation and assuming a spatially flat FLRW model. In order to infer $r_d$, the $r_d$-$H_0$ degeneracy is broken by adopting the SH0ES $H_0$ measurement of Ref.~\cite{Riess:2021jrx}, as discussed in Sec.~\ref{subsec:soundhorizon}. With our baseline settings, we infer $\beta_{\text{BAO}}=29.90 \pm 0.33$ and $r_d=136.20^{+2.20}_{-2.40}\,{\text{Mpc}}$. The orange band on the left side indicates the value of $\beta_{\text{BAO}}=30.10^{+0.22}_{-0.21}$ inferred within the reference $\Lambda$CDM model, whereas the value of $r_d=(147.21 \pm 0.23)\,{\text{Mpc}}$ inferred within the same model (without SH0ES calibration) is indicated by the green band on the right side.}
\label{fig:hubbletension}
\end{figure*}

\subsection{Impact of spatial curvature}
\label{subsec:omegak}

We now allow for a nonzero spatial curvature parameter $\Omega_K$. We stick once more to the cubic spline interpolation reconstruction, while adopting the DESI BAO and PantheonPlus SNeIa datasets. The results are shown in the right panel of Fig.~\ref{fig:sneiacurvature}, again with the same subfigure structure as in the left panel of Fig.~\ref{fig:splinevsgpdesivssdss}.

In this case, we find a very slight hint for negative spatial curvature (positive $\Omega_K$, corresponding to a spatially open Universe), with $\Omega_K = 0.135 \pm 0.087$.~\footnote{We note that this goes in the opposite direction compared to the possible preference for a spatially closed Universe from \textit{Planck} data~\cite{DiValentino:2019qzk}, see for instance Refs.~\cite{Handley:2019tkm,Efstathiou:2020wem,DiValentino:2020hov,Benisty:2020otr,Vagnozzi:2020rcz,Vagnozzi:2020dfn,DiValentino:2020kpf,Yang:2021hxg,Cao:2021ldv,Gonzalez:2021ojp,Dinda:2021ffa,Zuckerman:2021kgm,Bargiacchi:2021hdp,Glanville:2022xes,Bel:2022iuf,Yang:2022kho,Stevens:2022evv,Favale:2023lnp,Qi:2023oxv,Giare:2023ejv} for recent discussions.} The main effect of $\Omega_K>0$ is to allow for a higher $H(z)$ while at the same time also raising the corresponding distances, particularly at higher redshift. The net result is a slightly improved fit to the LRG1 $D_H/r_d$ point, while simultaneously improving the fit to the Ly-$\alpha$ line-of-sight $D_H/r_d$ and transverse $D_M/r_d$ points at $z_{\text{eff}}=2.330$, as is clear from the upper and middle subfigures in the right panel of Fig.~\ref{fig:sneiacurvature}. At the same time, the higher expansion rate allows for a slightly better fit to the bump feature at $z \sim 0.5$.

Overall, however, we notice that the inclusion of spatial curvature does not drastically alter the significance of the bump and depression features identified in the spatially flat case (if anything, these features are very slightly enhanced). Finally, for completeness we examine the impact of utilizing a different SNeIa dataset. Adopting the Union3 and DESY5 samples in place of the PantheonPlus catalog, we find that the previous (already very weak) hint for negative spatial curvature weakens significantly: specifically, we infer $\Omega_K = 0.098 \pm 0.089$ and $\Omega_K = 0.067 \pm 0.089$ respectively.

\subsection{Sound horizon and the Hubble tension}
\label{subsec:soundhorizon}

It is worth recalling that, in obtaining all the earlier results, we also varied $\beta_{\text{BAO}}=c/r_dH_0$, the dimensionless parameter which sets the overall scale of the observables relevant for BAO measurements -- $D_V/r_d$, $D_M/r_d$, and $D_A/r_d$. A visual summary of the values of $\beta_{\text{BAO}}$ we obtained for all variations over our baseline analysis is provided on the left side of Fig.~\ref{fig:hubbletension}. We see that all obtained values of $\beta_{\text{BAO}}$ are consistent, well within $1\sigma$, with the value obtained from a $\Lambda$CDM fit to \textit{Planck} and BAO data from BOSS DR12, MGS, and 6dFGS data ($\beta_\text{BAO} = 30.10^{+0.22}_{-0.21}$, right column of Table~2 in Ref.~\cite{Planck:2018vyg}), shown as the orange band. For reference, from our baseline cubic spline reconstruction combining DESI and PantheonPlus data, we infer $\beta_\text{BAO} = 29.90 \pm 0.33$.

The dataset combination we have considered is inherently uncalibrated. Calibrating it through an external constraint/prior on $r_d$ essentially amounts to constructing an inverse distance ladder from which we can estimate $H_0$. Conversely, calibrating it through an external constraint/prior on $H_0$ allows us to infer $r_d$. For concreteness and for illustrative purposes, we adopt the widely used SH0ES local distance ladder measurement of $H_0 = (73.04 \pm 1.04)\,\text{km/s/Mpc}$ from Ref.~\cite{Riess:2021jrx}. The resulting values of $r_d$ are shown on the right side of Fig.~\ref{fig:hubbletension}, with the value $r_d = 147.21 \pm 0.23\,{\text{Mpc}}$ obtained from a $\Lambda$CDM fit to \textit{Planck} and BAO data from BOSS DR12, MGS, and 6dFGS shown as the green band.

We see that the $\Lambda$CDM-based inference of $r_d$ is in strong tension with all the values of $r_d$ we obtained. For instance, in our baseline analysis this tension is at the $5\sigma$ level, increasing to $6\sigma$ when the DESI BAO dataset is replaced by the older SDSS BAO one. The fact that the value of $\beta_{\text{BAO}}$ decreases slightly when adopting the DESI BAO measurements in place of the SDSS ones (or equivalently that the corresponding value of $r_d$ increases once the value of $H_0$ used to calibrate the ladder is fixed) is in agreement with the value of $H_0$ obtained by the DESI collaboration being slightly higher than previous inferences based on earlier BAO data~\cite{DESI:2024mwx}. To put it differently, our nonparametric results are in agreement with the statement that DESI BAO data \textit{slightly} reduce the tension between local distance ladder and inverse distance ladder inferences of $H_0$, as explicitly discussed in Sec.~6 of the DESI cosmological parameters paper~\cite{DESI:2024mwx}. However, the right panel of Fig.~\ref{fig:hubbletension} shows that the ``sound horizon tension'' overall persists at very high significance, even when accounting for all the different analysis variations considered. It is also worth comparing the right panel of Fig.~\ref{fig:hubbletension} to the older and analogous Fig.~12 in Ref.~\cite{Bernal:2016gxb} and Fig.~3 in Ref.~\cite{Aylor:2018drw}: what was barely a $3\sigma$ tension in $r_d$ back then (also inferred in a nonparametric way), has now grown to the $5\sigma$-$6\sigma$ level.

\begin{figure*}[!t]
\centering
\includegraphics[width=0.5\linewidth]{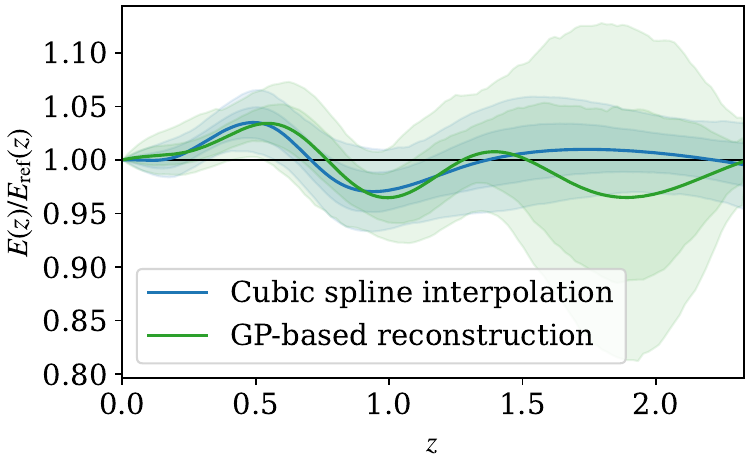}
\caption{Reconstructed posterior for the unnormalized expansion rate $E(z)$ relative to the reference $\Lambda$CDM model, obtained through cubic spline interpolation (blue) and Gaussian Process-based reconstruction (green). The dark and light bands indicate 68\% and 95\% credible intervals for the reconstructed function. We see that, overall, the shape of the reconstructed expansion history is constrained to deviate by no more than $\approx 10\%$ from the reference model.}
\label{fig:ez}
\end{figure*}

Taken at face value, the persistence of the resulting ``sound horizon tension'' confirms, if ever there was any need, that smooth modifications to the late-time expansion history cannot on their own solve the Hubble tension, as extensively argued in earlier works~\cite{Bernal:2016gxb,Addison:2017fdm,Lemos:2018smw,Aylor:2018drw,Schoneberg:2019wmt,Knox:2019rjx,Arendse:2019hev,Efstathiou:2021ocp,Cai:2021weh,Keeley:2022ojz}. In fact, it is worth stressing that our reconstructions are independent of the specific details of the late-time expansion history, provided the latter is sufficiently smooth. Therefore, pre-recombination new physics which reduces $r_d$ is absolutely needed (although, as argued elsewhere, this might on its own not be sufficient~\cite{Vagnozzi:2023nrq,Poulin:2024ken,Pedrotti:2024kpn}).

\section{Conclusions}
\label{sec:conclusions}

The era of Stage IV cosmology has now officially started thanks to the release of the first set of BAO measurements from DESI. The latter feature a potentially puzzling preference for an evolving dark energy component which, if substantiated, would have tremendous repercussions for what concerns our understanding of the Universe. Prompted by these puzzling features and the persisting Hubble tension, our goal in this work has been to reconstruct the late-time ($z \lesssim 2.33$) expansion history in a way which is as model-independent and nonparametric as possible. Under the minimal assumptions of a (not necessarily spatially flat) FLRW Universe, the validity of the Etherington distance-duality relation, and a smooth expansion history, we have considered two methods, whose results we verify are in excellent agreement with each other: an interpolation through natural piece-wise cubic splines (following Refs.~\cite{Bernal:2016gxb,Aylor:2018drw}), and a Gaussian process-based reconstruction (following Ref.~\cite{Ye:2023zel}). We have adopted DESI and SDSS data on the BAO side, and SNeIa data from the PantheonPlus, Union3, and DESY5 samples.

For what concerns the baseline reconstruction from DESI+PantheonPlus, both methods identify two features in the unnormalized expansion rate $E(z)$, with significance of $\gtrsim 2\sigma$ relative to a reference $\Lambda$CDM model: a bump at $z \sim 0.5$, and a depression at $z \sim 0.9$ (see the left panel of Fig.~\ref{fig:splinevsgpdesivssdss}). We find that both features are absent when replacing the DESI BAO dataset with the older SDSS ones. Since the effective volume covered by DESI is still smaller than that of SDSS despite the larger number of unique redshifts, these features (and in particular the $z \sim 0.5$ one) may in principle simply be the result of an unlucky sample variance fluctuation, a possibility which will be clarified very soon with upcoming data releases from DESI. The depression feature is evident in both the DESI+PantheonPlus and DESI+Union3 reconstructions, but not in the DESI+DESY5 one: aside from upcoming data releases from DESI, the origin of this feature may therefore be clarified by future SNeIa samples in the $z \lesssim 1$ range. We find that the $w_0w_a$CDM fit to DESI data is partially capturing the bump feature but not the depression one, as the model does not possess sufficient degrees of freedom/complexity (see once more the left panel of Fig.~\ref{fig:splinevsgpdesivssdss}). In fact, fitting both features requires an oscillatory/nonmonotonic behaviour. We have verified that allowing for a nonzero spatial curvature parameter does not appreciably alter any of the previous conclusions. Our inference of $c/r_dH_0 = 29.90 \pm 0.33$ confirms that, despite the DESI data allowing for more freedom in the late-time expansion history, the Hubble tension most certainly still requires pre-recombination new physics reducing the sound horizon (see the right panel of Fig.~\ref{fig:hubbletension}). Our nonparametric reconstruction of the unnormalized expansion rate $E(z)$ shown in Fig.~\ref{fig:ez} indicates that deviations in the $z \lesssim 2$ shape of the expansion history, relative to the reference $\Lambda$CDM model, are limited to $\lesssim 10\%$ at best: importantly, these limits apply to any smooth late-time modification to $\Lambda$CDM invoked to alleviate the Hubble tension -- or, as it might be more appropriate to refer to it, the ``cosmic calibration tension''~\cite{Poulin:2024ken}.~\footnote{See for instance Refs.~\cite{Zhao:2017urm,Vagnozzi:2018jhn,Yang:2018euj,Banihashemi:2018oxo,Banihashemi:2018has,Li:2019yem,Yang:2019nhz,Vagnozzi:2019ezj,Visinelli:2019qqu,DiValentino:2019jae,DiValentino:2019ffd,Hogg:2020rdp,Alestas:2020mvb,Banerjee:2020xcn,DiValentino:2020kha,Banihashemi:2020wtb,Alestas:2020zol,Gao:2021xnk,Alestas:2021xes,Heisenberg:2022lob,Heisenberg:2022gqk,Sharma:2022ifr,Nunes:2022bhn,Sharma:2022oxh,Reeves:2022aoi,Moshafi:2022mva,Banerjee:2022ynv,Alvarez:2022wef,Gangopadhyay:2022bsh,Gao:2022ahg,Dahmani:2023bsb,deCruzPerez:2023wzd,Ballardini:2023mzm,Yao:2023ybs,Gangopadhyay:2023nli,Zhai:2023yny,SolaPeracaula:2023swx,Gomez-Valent:2023hov,Frion:2023xwq,Petronikolou:2023cwu,Ben-Dayan:2023htq,daCosta:2023mow,Lazkoz:2023oqc,Forconi:2023hsj,Sebastianutti:2023dbt,Giare:2024ytc,Shah:2024rme,Giare:2024akf,Montani:2024pou,Aboubrahim:2024spa} for examples of late-time new physics models which can partially alleviate the Hubble tension, many of which involve new DE physics.}

Overall, the hints for potential late-time deviations from $\Lambda$CDM we have nonparametrically identified are intriguing and corroborate the earlier DESI findings. Although we believe one should exercise caution at least until the release of the next set of results from the DESI collaboration, which will hopefully clarify whether these are truly hints or unlucky statistical fluctuations, it could be interesting to take the reconstructed shape of the expansion history shown in Fig.~\ref{fig:ez} at face value and try to develop simple parametric models which can reproduce it. Presumably, once fitted back to DESI data, such models should perform significantly better than both $\Lambda$CDM and $w_0w_a$CDM (see, however, Ref.~\cite{Wolf:2023uno}). We note that some of the oscillatory/nonmonotonic features we identified bear partial resemblance to the so-called omnipotent DE model~\cite{DiValentino:2020naf,Adil:2023exv}, which could therefore be an interesting starting point in this sense.~\footnote{It has recently been argued that oscillatory features may be an inevitable outcome in models which help address the Hubble tension while featuring late-time modifications to $\Lambda$CDM~\cite{Akarsu:2022lhx}.} We defer these issues to future work.

\begin{acknowledgments}
\noindent We acknowledge the use of high performance computing services provided by the International Centre for Theoretical Physics Asia-Pacific cluster. J.-Q.J.\ acknowledges support from the Joint PhD Training program of the University of Chinese Academy of Sciences. D.P., S.S.C., and S.V.\ acknowledge support from the Istituto Nazionale di Fisica Nucleare (INFN) through the Commissione Scientifica Nazionale 4 (CSN4) Iniziativa Specifica ``Quantum Fields in Gravity, Cosmology and Black Holes'' (FLAG). S.S.C.\ acknowledges support from the Fondazione Cassa di Risparmio di Trento e Rovereto (CARITRO Foundation) through a Caritro Fellowship (project ``Inflation and dark sector physics in light of next-generation cosmological surveys''). S.V. acknowledges support from the University of Trento and the Provincia Autonoma di Trento (PAT, Autonomous Province of Trento) through the UniTrento Internal Call for Research 2023 grant ``Searching for Dark Energy off the beaten track'' (DARKTRACK, grant agreement no.\ E63C22000500003). This publication is based upon work from the COST Action CA21136 ``Addressing observational tensions in cosmology with systematics and fundamental physics'' (CosmoVerse), supported by COST (European Cooperation in Science and Technology).
\end{acknowledgments}

\appendix

\section{Impact of knots in cubic spline interpolation}
\label{appendix:knots}

\begin{figure*}[!ht]
\centering
\includegraphics[width=0.43\linewidth]{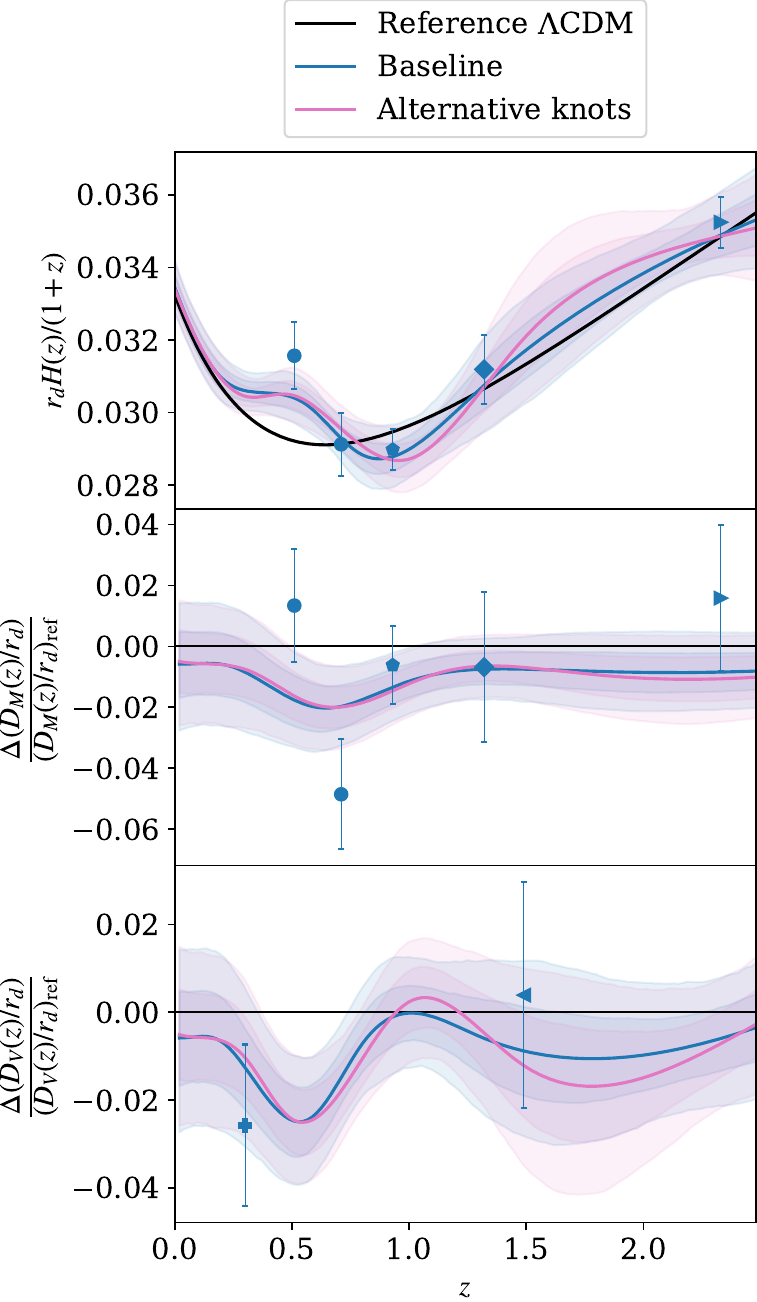}\,\,\,\,\,\,\,\,\,
\includegraphics[width=0.43\linewidth]{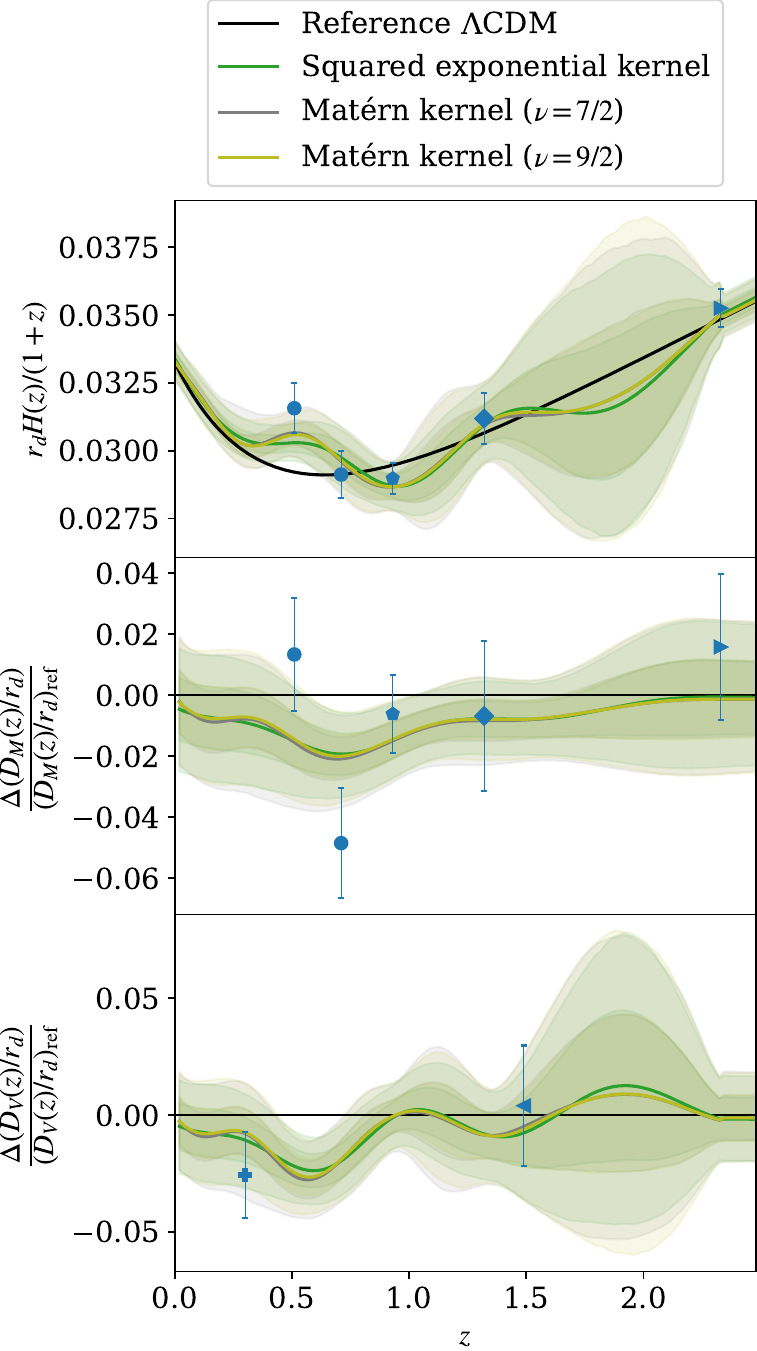}
\caption{\textit{Left panel}: comparison between the results obtained using the baseline set of knots for cubic spline interpolation (blue curves and bands), and with the alternative choice of knots presented in Appendix~\ref{appendix:knots} (magenta curves and bands), with the same subfigure structure as the left panel of Fig.~\ref{fig:splinevsgpdesivssdss}. In all cases, the DESI+PantheonPlus dataset combination has been adopted. \textit{Right panel}: comparison between the results obtained from the Gaussian-Process based reconstruction adopting the squared exponential kernel (green curves and bands), the Mat\'{e}rn-$7/2$ kernel (grey curves and bands), and the Mat\'{e}rn-$9/2$ kernel (gold curves and bands), as discussed in Appendix~\ref{appendix:kernel}, with the same subfigure structure as the left panel. In all cases, the DESI+PantheonPlus dataset combination has been adopted.}
\label{fig:knotskernel}
\end{figure*}

In the baseline reconstruction procedure based on cubic spline interpolation, we set our knots at $z_i=\{0.2, 0.57, 0.8, 1.3, 2.33\}$. To assess the impact of knot distribution, we consider a different choice of knots at $z_i=\{0.3, 0.5, 1.0, 1.5, 2.33\}$. A comparison between the results obtained from the two different choices of knots is shown in the left panel of Fig.~\ref{fig:knotskernel}. We see that the reconstructed functions agree well in the two cases, especially in redshift ranges where several data points are present. With the alternative choice of knots, we infer $\beta_{\text{BAO}}=29.94 \pm 0.32$, which is in excellent agreement with the baseline value of $\beta_{\text{BAO}}=29.90 \pm 0.33$.

\section{Impact of kernel in Gaussian Process-based reconstruction}
\label{appendix:kernel}

Our baseline Gaussian Process-based reconstruction adopted the squared exponential kernel, in light of its many advantageous features, which we extensively discussed in Sec.~\ref{subsubsec:gpbased}. Nevertheless, to assess the impact of the chosen kernel, we adopt two other choices of kernel widely used in cosmology. Both belong to the Mat\'{e}rn family, whose kernels are characterized by an index $\nu$ and take the following form:
\begin{eqnarray}
k(x_i, x_j) &=&\sigma^2 \frac{1}{\Gamma(\nu)2^{\nu-1}} \left ( \frac{\sqrt{2\nu}}{\ell} \vert x_i - x_j \vert \right )^\nu \,, \nonumber \\
&\times &K_\nu \left ( \frac{\sqrt{2\nu}}{\ell} \vert x_i - x_j \vert \right ) \,,
\label{eq:matern}
\end{eqnarray}
where $\Gamma$ is the gamma function, $K_{\nu}$ is the modified Bessel function of $\nu$th order, and $\nu>0$ controls the smoothness of the function. In the $\nu \to \infty$ limit, the Mat\'{e}rn kernel converges to the squared exponential one. We consider two cases, namely $\nu=7/2$ and $\nu=9/2$. The resulting reconstructed functions are shown in the right panel of Fig.~\ref{fig:knotskernel}. We see that the differences between the squared exponential, Mat\'{e}rn-$7/2$, and Mat\'{e}rn-$9/2$ reconstructions are for all intents and purposes negligible, confirming the stability of our GP-based reconstruction results against the chosen kernel. The only noteworthy feature is that the significance of the $z \approx 0.5$ bump is slightly enhanced when adopting either of the Mat\'{e}rn kernels: this is not surprising, as these kernels are known to be better suited at capturing local features with respect to the squared exponential one.

\clearpage

\bibliography{ezdesi}
\end{document}